%% file: paper.tex
\documentclass{article}
\usepackage{spconf,amsmath,graphicx}
\usepackage{booktabs}
\usepackage{makecell}
\usepackage{multirow}
\usepackage{tikz}
\usepackage{url}
\usepackage{amssymb}
\usepackage{pifont}% http://ctan.org/pkg/pifont

\usepackage{tabu}
\usepackage{tabularx} % in the preamble

\usepackage{subfig}
\usepackage{listings}
\usepackage{color}

\definecolor{dkgreen}{rgb}{0,0.6,0}
\definecolor{gray}{rgb}{0.5,0.5,0.5}
\definecolor{mauve}{rgb}{0.58,0,0.82}

\title{Streaming Multi-Speaker ASR with RNN-T}
\name{Ilya Sklyar*, Anna Piunova*, Yulan Liu\thanks{*These authors have contributed equally.}}
\address{Amazon Alexa\\
	\small \ttfamily ilsklyar@amazon.com\hspace{1cm}piunova@amazon.com\hspace{1cm}lyulan@amazon.com
}

\begin{document}
%\ninept
%
\maketitle

\input{abstract}
\input{introduction}
\input{technical_approach}
\input{data_and_experimental_setup}

\input{results_and_discussion}
\input{conclusion}
\input{acknowledgement}

\vfill
\pagebreak
%\section{REFERENCES}
%\label{sec:refs}
%
%List and number all bibliographical references at the end of the
%paper. The references can be numbered in alphabetic order or in
%order of appearance in the document. When referring to them in
%the text, type the corresponding reference number in square
%brackets as shown at the end of this sentence \cite{C2}. An
%additional final page (the fifth page, in most cases) is
%allowed, but must contain only references to the prior
%literature.

%\bibliographystyle{IEEEbib}
\bibliographystyle{IEEEtran}
\bibliography{strings,refs}
%\vfill
%\pagebreak
%\input{appendices}
\end{document}

%% file: abstract.tex
\begin{abstract}
Recent research shows end-to-end ASR systems can recognize 
overlapped speech from multiple speakers. 
However, all published works have assumed no latency constraints during inference,
which does not hold for most voice assistant interactions. This work focuses on 
multi-speaker speech recognition based on a recurrent neural network transducer (RNN-T) that has been 
shown to provide high recognition accuracy at a low latency online recognition regime. 
We investigate two approaches to multi-speaker model training of the RNN-T: deterministic output-target assignment and permutation invariant training. 
We show that guiding separation with speaker order labels in the former case enhances the high-level speaker tracking capability of RNN-T.
Apart from that, with multi-style training on single- and multi-speaker utterances, the resulting models gain robustness against ambiguous 
numbers of speakers during inference. Our best model achieves a WER of 10.2\% on simulated 2-speaker LibriSpeech data, which is competitive with the previously reported state-of-the-art non-streaming model (10.3\%), 
while the proposed model could be directly applied for streaming applications.
\end{abstract}

%% file: introduction.tex
\section{Introduction}
\label{sec:introduction}
Significant progress has been made lately in recognizing partly or fully overlapped 
speech from multiple speakers in single-channel audio recordings.
Research work in this field can be grouped into two algorithmic families. 
One family of algorithms \cite{Isik_2016, Menne_2019, Settle_2018, von_Neumann_2020} 
represents modular systems with dedicated speaker separation front-ends 
such as deep clustering \cite{Hershey_2016} or TASnet \cite{Luo_2018} that 
are trained or pre-trained with signal reconstruction objective, 
thus requiring parallel clean data without overlapped speech. 
Another family of methods \cite{Yu_2017_icassp, Yu_2017_interspeech, Qian_2018, Seki_2018, Chang_2019} 
trains multi-speaker ASR models against the target multi-speaker transcriptions 
directly, without intermediate signal reconstruction to separate the speech per 
speaker. These systems typically rely on permutation-invariant training (PIT) to 
consider all possible pairs between model outputs and reference transcriptions. 
Such models can be trained without original clean signals, however, their scalability 
is limited by the hard restriction on the maximum number of speakers ($S$) and 
the high computation complexity of the PIT algorithm ($S!$). 
A recently proposed method, serialized output training (SOT)
\cite{k2020serialized, kanda2020joint}, tackles these limitations of PIT by ordering 
the recognition output for different speakers sequentially, i.e. one speaker after 
another with speech transcription separated by a speaker change tag.

Many voice assistant applications operate in a streaming mode for
voice-activated two-way interactions between user and device, thus maintaining
low latency is as important as achieving high accuracy. 
%The modular systems with stacked components are 
%disadvantaged to maintain low latency compared to the highly integrated 
%end-to-end (E2E) ASR systems that directly convert audio to transcription. 
The recent work on SOT 
\cite{k2020serialized, kanda2020joint} is based on a type of E2E models which is not optimized for a low latency setup as the attention module assumes
that the full input mixture is available. Besides, SOT by design introduces 
delays in recognizing the speech from non-leading speakers, thus
fundamentally incompatible with streaming applications.

Recent research in \cite{shafey2019} shows that low latency E2E models such as
recurrent neural network transducer (RNN-T \cite{graves2012sequence}) can recognize speech from multiple speakers and tag speakers based on their roles. This motivates us to propose a modified RNN-T model to process overlapping audio in a time-synchronous fashion and emit text sequences for multiple speakers in parallel. 
While previous work \cite{shafey2019, Tripathi_2020} experimented with RNN-T in 
multi-talker scenario, \cite{shafey2019} did not consider the cross-talk between 
speakers and \cite{Tripathi_2020} relied on a non-streaming bi-directional LSTM 
encoder in their RNN-T model. Therefore, the present work is the first attempt to build a streaming multi-speaker ASR system with RNN-T to the best of our knowledge.

We revisit two previously studied approaches to multi-speaker ASR model training in the RNN-T setting: 
deterministic output-target assignment and PIT. 
For the former, we explore informing
RNN-T with speaker order labels to strengthen its capability in speaker
tracking.
We also study using multi-style training to improve generalization to an uncertain number of speakers during inference.
The evaluation setup is shared with \cite{k2020serialized,kanda2020joint} 
and the best performance we achieved on 2-speaker partially overlapping mixtures
is competitive with non-streaming multi-speaker ASR systems. 

%% file: technical_approach.tex
\section{Technical approach}
\label{sec:technical-approach}
\subsection{RNN-T model}
\label{ap:tech-approach}

\begin{figure*}[t]
	\centering
	\subfloat[\label{(a)}]{\includegraphics[width=0.35\linewidth]{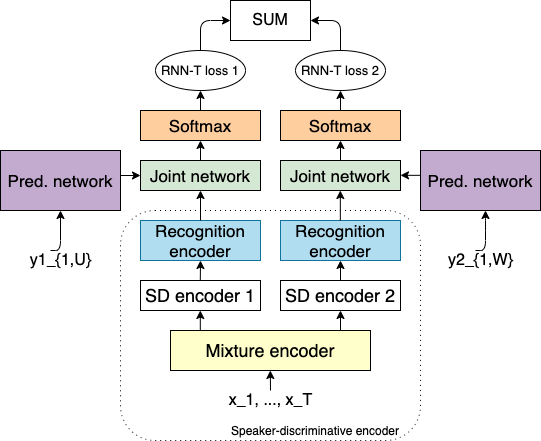}}
	\hfil
	\centering
	\subfloat[\label{(b)}]{\includegraphics[width=0.35\linewidth]{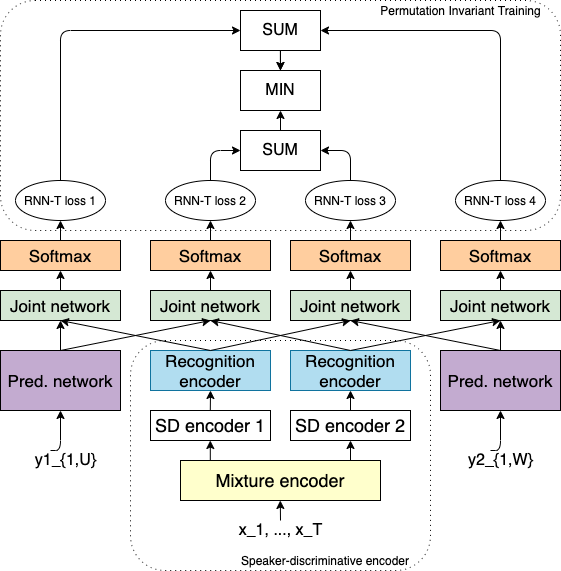}}
	\caption{Experimental multi-speaker RNN-T models. (a) DAT-MS-RNN-T: deterministic assignment between outputs of the encoder and prediction network (in the order of appearance). (b) PIT-MS-RNN-T: optimal assignment with permutation invariant training. Blocks with the same colour share parameters between speakers.}
	\label{fig:ms-rnnt}
\end{figure*}

RNN-T \cite{graves2012sequence} model is a popular 
choice for streaming ASR applications with the state-of-the-art recognition 
performance. Given a sequence of acoustic feature vectors 
$\mathbf{x}$ and the corresponding label sequence % $\mathbf{x} = (x_1, \ldots, x_T)$
$\mathbf{y}$, the model estimates a conditional probability % $\mathbf{y} = (y_1, \ldots, y_U)$
distribution $P(\mathbf{y}| \mathbf{x})$.

A standard single speaker RNN-T consists of three main building blocks: 
an encoder, a prediction network and a joint network.
The encoder sequentially processes acoustic features 
and generates high-level representations 
$\mathbf{h} = \operatorname{Enc}(\mathbf{x})$,
%$\textbf{H}^{enc} = (h_{1}^{enc}, ..., h_{T}^{enc})$, 
similarly to acoustic model in hybrid ASR. The prediction network models the next 
label \(y_{j+1}\) given previous labels in the 
sequence \( (y_1, \ldots, y_j)\). During training, ground truth is used as a previous label context in
prediction network input, while during inference previous non-blank prediction output is used. The joint network is a feed-forward network that processes the output from both encoder and prediction network and outputs the probability distribution over all output labels.

Loss function optimized during RNN-T model training is defined as a negative log-likelihood: 
\begin{equation}
\label{eq:rnnt-objective}
\mathcal{L} = -  \log P(\mathbf{y}| \mathbf{h}),
\end{equation}
where \( P(\mathbf{y}  | \mathbf{h}) = \sum_\mathbf{\hat{y}} P(\mathbf{\hat{y}} | \mathbf{h}) \), \( \mathbf{\hat{y}}  \in \mathcal{A}\). \(\mathcal{A}\) is the set of all the possible alignments (with blank labels) between encoder features \( \mathbf{h}\) and targer label sequence \( \mathbf{y} \).

\subsection{Multi-speaker RNN-T model}

This work extends the standard single speaker RNN-T to a multi-speaker RNN-T capable to recognize
speech of each speaker in the presence of overlapped speech from 
different speakers.
In the multi-speaker RNN-T (MS-RNN-T) model, speech separation is performed within the encoder. This strategy follows the approach initially proposed in \cite{Seki_2018} for the joint 
CTC/attention-based encoder-decoder network \cite{kim2017joint,hori-etal-2017-joint}. 
At the same time, the prediction network and joint network topologies remain unchanged, 
and their parameters are shared across speakers.
% to ensure the speaker-invariant behaviour of these modules.

As shown in Fig. \ref{fig:ms-rnnt}, the encoder of the multi-speaker RNN-T consists of a mixture encoder, speaker-dependent (SD) encoders
and a recognition encoder and is referred to as a speaker-discriminative encoder in the following. Input features are firstly processed by
the mixture encoder which extracts acoustic representation of the mixed speech. 
Mixture representation is fed into SD encoders which have different parameters per speaker, hence enforcing the model to generate separated intermediate encoding representations. 
This design implicitly assumes that the maximum possible number of cross-talking speakers 
$S$ is known in advance. The speaker-dependent representations are 
processed in parallel by the recognition encoder module, whose parameters are shared 
across all speakers. Mathematical formulation of the mixed speech processing for each speaker \(s \in (1,...,S) \) is presented below:
\vspace*{2mm}
\begin{align}
	\mathbf{h}^{MixEnc} &= \operatorname{MixEnc}(\mathbf{x}) \label{eq:mixture-encoder} \\
	\mathbf{h}^{SDEnc}_s &= \operatorname{SDEnc_s}(\mathbf{h}^{MixEnc}) \label{eq:sp-encoder} \\
	\mathbf{h}_s &= \operatorname{RecEnc}(\mathbf{h}^{SDEnc}_s) \label{eq:recognition-encoder}
\end{align}

To compute alignments between separated encoder features \( \mathbf{h}_s \) and corresponding target label sequence \( \mathbf{y}_s \) we need to find an assignment between outputs of the speaker-discriminative encoder and the prediction networks. Speaker assignment is challenging in such a multi-output architecture, as it is unclear which hypothesis-reference speaker pairs to use for training. One simple approach is to force the model to implicitly learn speaking order via deterministic assignment training (DAT) similarly to  \cite{k2020serialized, kanda2020joint, Tripathi_2020}. Considering the scenario of two speakers, the model with two outputs can be trained in a way so that output 1 is always optimized towards label sequence for the speaker who started speaking first (speaker 1), while output 2 is optimized towards the label sequence for the follow-up speaker (speaker 2), as shown in Fig. \ref{fig:ms-rnnt} (a). Following \cite{Tripathi_2020}, we facilitate correct assignment between model outputs and targets during training by adding a fixed speaker order label to each input of the corresponding SD encoder. The total loss of DAT-MS-RNN-T is a sum of the standard RNN-T losses for all speakers:

\begin{equation}
\begin{split}
\mathcal{L} = - \sum_{s} \log P(\mathbf{y}_{s} | \mathbf{h}_s )
\label{eq:multi-sp-loss}
\end{split}
\end{equation}

As an alternative to DAT, permutation-invariant training can be used to provide an 
additional degree of freedom to pick the optimal assignment. Following this training 
approach, model parameters are optimized with respect to the best speaker 
assignment pairs that are chosen by evaluating all possible assignments and selecting 
the one with the lowest overall loss. Standard RNN-T objective function is combined 
with PIT in PIT-MS-RNN-T to optimize the joint loss based on the best permutation, as depicted in 
Fig. \ref{fig:ms-rnnt} (b) and described as follows:

\begin{equation}
\begin{split}
\mathcal{L}_{PIT} =  \min \limits_{\pi \in \mathcal{P}} - \sum_{s} \log P(\textbf{y}_{s} | \mathbf{h}_{\pi(s)} ),
\label{eq:multi-sp-loss-pit}
\end{split}
\end{equation}
where \( \pi(s)  \) defines a permutation \(\pi\) of speaker \( s \) from the set of permutations \( \mathcal{P} \). In the considered scenario of two-speaker overlap, \( \mathcal{P} = \{(1,2), (2,1)\} \)

%% file: data_and_experimental_setup.tex
\section{Data and experimental setup}
\vspace*{-1mm}
\label{sec:data-and-experimental-setup}
\subsection{Data}
\label{sec:data}
\vspace*{-1mm}
Experiments are conducted on simulated data based on LibriSpeech corpus \cite{Panayotov_libri}.
The simulation design and mixture configurations are replicated from \cite{k2020serialized} with the help from
authors. The training dataset consists of 281241 utterances (around 1.5k hours) with a partially overlapping speech from 2 speakers, where each speaker  turns in once per utterance. Delay for second speaker is randomly sampled with two constraints: 1) at least 0.5 sec 
time difference between the speech start of two speakers; 
2) each mixture contains an overlapping segment.
The dev and test partitions are generated from utterances in dev-clean and test-clean partitions of LibriSpeech dataset, respectively, with relaxed constraint 1). The resulting dataset has an overlap ratio of 28\%, 25\% and 24\% for train, dev and test partition respectively.
\vspace*{-1mm}
\subsection{Evaluation metric}
\label{sec:evaluation-metric}
\vspace*{-1mm}
In multi-speaker setup, a model might produce more than one output sequences. For the given input audio stream, the multi-speaker model generates 
$S$ outputs that corresponds to the number of overlapping speakers.
% In the presence of two speakers in the mixture, we have two reference transcriptions $T1$ and $T2$. However, we don’t know which of the two generated results $R1$ and $R2$ should be compared to which of the ground truth sequences. 
%In this work we propose different strategies for WER calculation to address the
%ambiguity in speaker assignment between hypothesis and reference. 
Consistent to \cite{k2020serialized, Tripathi_2020} we calculate edit distances for all possible permutations between outputs and targets and  
choose pairs with the minimal sum. The overall WER is thus 
the optimal sum of edit distances normalized by 
the sum of reference lengths of each target transcription.
Scoring is consistent across all reported results, including DAT models.
Baseline single-speaker RNN-T generates one output sequence by design.
When evaluating it's performance on mixed speech we hypothesize that 
there are other empty outputs and treat their edit distance as deletions.
The corresponding WER is calculated as described before.
%Alternatively, we also calculate the WER using 
%deterministic assignment, where model output 1 is always evaluated against 
%speaker 1, while model output 2  is evaluated against speaker 2. 
%Such WERs could indicate how well the model can implicitly track 
%speaker in the correct speaking order.
%TODO: elaborate on the metric, add equations
\vspace*{-3mm}
\subsection{Experimental setup}
\label{sec:experimental-setup}
\vspace*{-2mm}
Multi-speaker models are based on the architecture proposed in
Section \ref{sec:technical-approach}. We use two-layer LSTMs with 1,024 units in each module of the speaker-discriminative encoder. 
The prediction network is a two-layer LSTM with 1,024 units in each layer. 
The output size of the recognition encoder and the prediction network is set to 
640. We use a one-layer feed-forward joint network with 512 units. The output softmax layer dimensionality is 2501 which corresponds to blank label + 2500 word pieces: the most likely subword segmentation from a 
unigram word piece model \cite{kudo2018sentencepiece}. Acoustic features are 64-dimensional log-mel filterbanks with a frame shift of 10ms which are stacked and downsampled by a factor of 3.
For feature augmentation we employ LibriFullAdapt SpecAugment policy from 
\cite{Park_2020}. We use Adam algorithm \cite{kingma2014adam} for optimization of all models, and the learning rate is scheduled based on warm-up, hold and decay strategy as proposed in \cite{Park_2019}. For each experimental run, the best model is chosen based on its performance on the development set.
%There is no LM rescoring or shallow fusion involved for the sake of simplicity. %

Our single-speaker baseline has the same architecture as multi-speaker RNN-T with only one active speaker-discriminative encoder branch. 
We train it on 960 hours of single-speaker LibriSpeech data and employ it as a seed model for all subsequent experiments on multi-speaker data. 
%For \emph{SD encoder 2} initialization of the multi-speaker RNN-T model we use two different approaches, depending on whether fixed speaker order labels are passed or not as input to speaker-dependent blocks of the encoder.
%In the former case we initialize \emph{SD encoder 2} with learned parameters of \emph{SD encoder 1}. 
%However, in the latter case inputs to both \emph{SD encoder 1} and \emph{SD encoder 2} are the same which causes collision at the start of fine-tuning on mixed data. 
%To ensure that model parameters are different enough we rely on the initialization approach proposed in \cite{Seki_2018} and add random noise to all weights of \emph{SD encoder 2}.

%% file: results_and_discussion.tex
\section{Results and discussion}
\label{sec:results-and-discussion}

%\begin{table}[th]
%	\vspace{-5pt}
%	\caption{Results with our single-speaker RNN-T model on mixed LibriSpeech \textit{test-2spk} partition}
%	\label{tab:baseline-2spk-result}
%	\centering
%	\begin{tabular}{ c c c c c}
%		\toprule
%		\multirow{2}{*}{Eval  type} & \multicolumn{4}{c}{Performance on test-2spk (\%)} \\ 
%		& WER & SUB & INS & DEL \\ 
%		\midrule
%		\makecell{w.r.t. each target}  &  98.6 & 28.6 & 67.73 & 2.19 \\
%		\midrule
%		Serialized &  35.69 & 14.42 & 20.17 & 1.92 \\
%		\midrule
%		\makecell{Oracle segmentation} &  37.61 & 26.32 & 7.37 & 3.92 \\
%		\bottomrule
%		\vspace{-10pt}
%	\end{tabular}
%	
%\end{table}

%TODO: add average WER among all test sets to table 1

\begin{table}[t]
	\vspace{-5pt}
	\caption{WERs [\%] on LibriSpeech test partitions and simulated 2-speaker test set achieved with different variants of the multi-speaker RNN-T model and single-speaker baseline.}
	\label{tab:multispeaker-2spk-model-tuning}
	\centering
	\begin{tabular}{ l c c c c }
		\toprule
		Model   & clean & other & 2spk & Overall \\
		\midrule
		RNN-T & 6.5 &  15.5 & 66.3 &  38.7 \\
		\midrule
		 DAT-MS-RNN-T  & 9.2 & 16.9 & 11.8 & 12.4\\
		 \hspace{3mm}+ speaker order label  & 7.7 & 16.2 & 11.7 & 11.8 \\
		 \hspace{6mm} +multi-style & 7.5 & 15.4 & 11.0 & 11.2 \\
		  PIT-MS-RNN-T & 7.9  & 15.8 & 10.6 & 11.2 \\
		  \hspace{3mm} +multi-style & 7.6 & 15.2 & 10.2 & 10.8 \\
		\midrule
		\vspace{-23pt}
	\end{tabular}
	
\end{table}

\subsection{Baseline results}
\label{sec:baseline-performance}
First, we evaluate the performance of the single-speaker baseline on LibriSpeech test partitions and simulated 2-speaker test mixtures. Results are presented in Table \ref{tab:multispeaker-2spk-model-tuning}. 
We observe severe WER degradation on the 2-speaker test set (\emph{test-2spk}) which is in line with other reports on this task achieved by non-streaming  ASR models \cite{k2020serialized}. 

\subsection{Results of multi-speaker RNN-T model variants}
\label{sec:multi-speaker-rnnt-results}
In the following set of experiments, we extend RNN-T model to 2-speaker processing, train it on simulated multi-speaker data and investigate how it generalizes to unseen mixtures as well as general single-speaker test utterances. 
We differentiate between two training approaches for MS-RNN-T models discussed in Section \ref{sec:technical-approach}: DAT and PIT. 

Surprisingly, simple DAT-MS-RNN-T model achieves a solid 82\% relative WER improvement over the 
single-speaker baseline on \textit{test-2spk}. 
At the same time, degradation on single-speaker utterances is observed.
Further analysis found that DAT-MS-RNN-T occasionally splits the hypotheses of a single speaker utterance in two outputs.
This indicates that such a multi-output model needs further improvement on speaker tracking to prevent hallucinating hypotheses from non-existing speakers.

Providing the SD encoders with speaker order labels helps the model to follow the same speaker better, resulting in 16\% and 4\% relative WER reduction reported on \emph{test-clean} and \emph{test-other} sets, respectively.
To bridge performance gap in the mismathched scenario further we combine single-speaker LibriSpeech dataset with simulated 2-speaker data and perform multi-style training on the pooled dataset. 
With this approach we observe relative WER gains of 3\% and 5\% on \emph{test-clean} and \emph{test-other} partitions, respectively.
Interestingly, recognition performance on 2-speaker utterances is also improved by 5\% relative, which could be explained by a larger variety of training data that improved generalization.

Our experiments with PIT-MS-RNN-T show that it outperforms DAT-MS-RNN-T with speaker order label by 9\% WER relative on \emph{test-2spk}.
With multi-style training, similar improvements are observed on PIT-MS-RNN-T as reported for DAT-MS-RNN-T. 
The overall WER achieved with PIT-MS-RNN-T model is 4\% relatively lower than the best DAT-MS-RNN-T model, with main performance improvements on overlapped speech.

\vspace*{-3mm}
\subsection{Comparison with attention-based models}
\vspace*{-1mm}
\label{sec:comparison-results}
In Table \ref{tab:multispeaker-2spk-result} our best streaming multi-speaker ASR model (PIT-MS-RNN-T) is benchmarked 
against non-streaming methods on \emph{test-2spk} and \emph{test-clean}. 
For non-streaming ASR models we consider attention-based encoder-decoder (AED) architectures: PIT-AED and SOT-AED from \cite{k2020serialized} and improved SOT-AED variants from \cite{kanda2020joint}.  It is important to underline that SOT-AED model with speaker identification uses an external data source for embedding extractor training.
Among all methods relying solely on LibriSpeech-based data, PIT-MS-RNN-T achieves better or comparable performance on the 2-speaker test set while being both light-weight and streaming-capable. Algorithmic latency of PIT-MS-RNN-T is equivalent to the feature frame rate of 30ms. Detected underperformance on \emph{test-clean} can be explained by the previously mentioned suboptimality of speaker tracking in multi-output models.
In \cite{kanda2020joint}, adding speaker information improved the accuracy on both single speaker and 2-speaker testsets.
Similar approaches can be applied on PIT-MS-RNN-T and DAT-MS-RNN-T in future work for applications where a speaker inventory is available.

\begin{table}[t]
	\vspace{-5pt}
	\caption{WERs [\%] comparison of the best streaming multi-speaker RNN-T model with non-streaming approaches on \textit{test-clean} and \textit{test-2spk} partitions}
	\label{tab:multispeaker-2spk-result}
	\centering
	\begin{tabular}{ c c c c c}
		\toprule
		Model & \makecell{\#params}   & \makecell{\#speakers \\ in training}  & clean & 2spk\\ 

		\midrule
		PIT-AED\cite{k2020serialized} & 160.7M   & 1,2    & 6.7 &  11.9 \\
		SOT-AED \cite{k2020serialized} & 135.6M   & 1,2,3   & 4.6 &  11.2 \\
		SOT-AED\cite{kanda2020joint} & 135.6M & 1,2,3   & 4.5 & 10.3 \\
		\hspace{3mm} + speakerID & 145.5M & 1,2,3 & 4.2 & 8.7 \\
		\midrule
		PIT-MS-RNN-T & 80.9M  &  1,2 & 7.6 & 10.2  \\
		\bottomrule
		\vspace{-18pt}
	\end{tabular}
	
\end{table}

%% file: conclusion.tex
\section{Conclusion}
\label{ap:conclusion}
\vspace*{-1mm}
We proposed a novel multi-speaker RNN-T model architecture which can be applied directly in streaming applications. 
We experimented with the proposed architecture in two different training scenarios: with deterministic and optimal assignment between model outputs and target transcriptions. 
We investigated the impact of explicit speaker order label conditioning and multi-style training on generalization to unseen single- and multi-speaker data.
Our best multi-speaker RNN-T model achieved on-par performance with non-streaming methods studied in the literature on the overlapped dataset with 2 speakers, which makes it a promising approach for further research in the field of a low-latency multi-speaker ASR.

%% file: acknowledgement.tex
\section{Acknowledgement}
\label{ap:acknowledgement}
\vspace*{-1mm}
We would like to acknowledge Naoyuki Kanda for openly sharing the details 
of their previous work \cite{k2020serialized,kanda2020joint}.
In addition, we would like to acknowledge the Alexa ASR teams for providing
both infrastructure and technical support that this work has heavily benefited from.